\documentstyle[prb,aps,multicol, epsfig]{revtex}
\begin{document}
\draft
\widetext
\title{Anisotropy of the Upper Critical Field and Critical Current in
Single Crystal
MgB$_2$}
\author{L.~Lyard}
\address{Laboratoire d'Etudes des Propri\'et\'es Electroniques des
Solides, Centre National de la Recherche Scientifique, BP166, F-38042 Grenoble
Cedex 9, France}
\address{Commissariat \`a l'Energie Atomique - Grenoble,
D\'epartement de Recherche Fondamentale sur la Mati\`ere Condens\'ee,
SPSMS, 17 rue des Martyrs, F-38054 Grenoble Cedex 9, France}
\author{P.~Samuely}
\address{University Joseph Fourier, BP 53, F-38041 Grenoble Cedex 9,
France}
\address{Institute of Experimental  Physics, Slovak Academy of
Sciences, SK-04353 Ko\v{s}ice, Slovakia}
\author{P.~Szabo}
\address{Institute of Experimental  Physics, Slovak Academy of
Sciences, SK-04353 Ko\v{s}ice, Slovakia}
\author{T.~Klein}
\address{Laboratoire d'Etudes des Propri\'et\'es Electroniques des
Solides, Centre National de la Recherche Scientifique, BP166, F-38042 Grenoble
Cedex 9, France}
\address{Institut Universitaire de France and University Joseph Fourier, BP 53, F-38041 Grenoble Cedex 9,France}
\author{C.~Marcenat, L. Paulius}
\address{Commissariat \`a l'Energie Atomique - Grenoble,
D\'epartement de Recherche Fondamentale sur la Mati\`ere Condens\'ee,
SPSMS, 17 rue des Martyrs, F-38054 Grenoble Cedex 9, France}
\author{K.~.~P.~Kim,  C.~U.~Jung, H.-S.~Lee, B.~Kang, S. Choi, S.-I. Lee}
\address{NVCRICS and Dept. of Physics, Pohang University of Science
and Technology, Pohang 790-784, Republic of Korea}
\author{J. Marcus, S.~Blanchard}
\address{Laboratoire d'Etudes des Propri\'et\'es Electroniques des
Solides, Centre National de la Recherche Scientifique, BP166, F-38042 Grenoble
Cedex 9, France}
\author{A.~G.~M.~Jansen}
\address{Grenoble    High     Magnetic    Field    Laboratory,
Max-Planck-Institut   f\"{u}r   Festk\"{o}rperforschung   and
Centre National de  la Recherche Scientifique,  B.P. 166, F-38042  Grenoble
Cedex 9, France}
\author{U.~Welp, G.~Karapetrov and W.~K.~Kwok}
\address{Materials Science Division, Argonne National Laboratory,
Argonne, Illinois 60439, USA}
\date{today}
\maketitle
\widetext
\begin{abstract} We  report on specific  heat, high magnetic
field  transport  and   $ac-$susceptibility  measurments  on
magnesium diboride single crystals. The upper critical field
$H_{c2}$ for  magnetic fields perpendicular  and parallel to
the Mg and  B planes is presented for the  first time in the
entire  temperature  range.  A  very  different  temperature
dependence  has been  observed in  the two  directions which
yields  to a  temperature dependent  anisotropy with $\Gamma
\sim$ 5 at  low temperatures and about 2  near $T_c$. A peak
effect is observed in  the susceptibility measurments for $\mu_0H
\sim $2 T parallel to  the $c-$axis and the critical current
density  presents a  sharp maximum  for $H$  parallel to the
$ab-$plane. \end{abstract}
\begin{multicols}{2} \narrowtext
\section{INTRODUCTION}     Since     the     discovery    of
superconductivity in  magnesium diboride at 39  K in January
2001 \cite{nagamatsu} enormous amount  of work has been done
which helped  to elucidate many of  its physical properties.
Among  others  the  multiband  electronic structure proposed
theoretically    \cite{liu,choi}     with    the    multiple
superconducting energy gaps has  been approved. The two main
superconducting energy gaps have already been experimentally
evidenced by different techniques like for instance specific
heat   measurments  \cite{bouquet}   or  Andreev  reflection
\cite{szabo}. A larger gap  is attributed to two-dimensional
$p_{x-y}$  orbitals and  a smaller  gap to three-dimensional
$p_z$  bonding  and  antibonding  orbitals.  Such  a picture
indicates  a significant  anisotropy of  the superconducting
state.  Data  reported  so  far  on  the  anisotropy  factor
$\Gamma =  H_{c2||ab}/H_{c2\bot ab}$ scatter from  1.1 to 13
depending on the form of material (polycrystals, thin films,
single  crystals)  and  method  of  evaluation.  The problem
remains to be clarified on high quality single crystals.\\
The paper is organized  as follows. The experimental details
are  given   in  section  II.   In  section  III,   we  show
a consistent  way  of  extracting  the  upper critical field
$H_{c2}$ for  magnetic fields parallel  and perpendicular to
the basal  $ab$ planes. $H_{c2}$ has  been determined in the
entire temperature range by high field magnetotransport, and
those  values   are  compared  to  the   ones  deduced  from
$ac$-susceptibility below 5 T  and specific heat measurments
below 7   T.  The  perpendicular   critical  field  reveals
a conventional   character   with   a   linear   temperature
dependence  near  $T_c$  and  $\mu_0H_{c2\bot  ab}(0) \simeq$ 3.5
T while   $H_{c2||ab}$   shows   a   positive  curvature  at
temperatures above $20$ K  with $\mu_0H_{c2||ab}(0) \simeq$ 17 T.
As   a  consequence   the  anisotropy   factor  $\Gamma$  is
temperature  dependent  with  $\Gamma$(0  K)  $\sim$  5  and
$\Gamma$(near $T_c$) about 2.  The angular dependence of the
upper  critical field  measured at  5.4 K  and 26  K show an
elliptic  form as  predicted by  a one-band  Ginzburg-Landau
theory, but  obviously this theory  can not account  for the
temperature  dependent  $\Gamma$   parameter.  Finally,  the
magnetic  field  and  angular  dependence  of  the  critical
current  deduced  from  $ac-$susceptibility  measurments  is
discussed in section IV.\\
\section{SAMPLE PREPARATION AND EXPERIMENT}
Experiments  have  been  performed  on  high quality MgB$_2$
single  crystals   coming  from  one   batch\cite{kim}.  The
crystals  show clear  hexagonal facets  with flat  and shiny
surfaces.  The typical  dimensions are  50x50x10 $\mu$m$^3$.
The resistivity has been measured at  13 Hz as a function of
magnetic  field up  to $28$  T for  different temperatures and
angles  ($\theta$) between  $H$ and  the $ab$-planes  of the
crystal.  The magnetic  field was  always orthogonal  to the
measuring current.  Gold electrodes have  been evaporated as
stripes overlapping the top plane of the sample with contact
resistance   of   about   1 $ \Omega$   in  the  four-probe
configuration. The $ac-$susceptibility  was deduced from the
local transmittivity  measured with a  miniature Hall probe.
The $ac$  excitation field ($h_{ac}  \sim 3$G, $\omega  \sim
23$Hz) was perpendicular to the  $ab-$planes and to the Hall
probe  plane  and  superimposed  to  a  $dc$ field which was
making  an angle  $0 <  \theta <  90$ with  the planes.  The
specific   heat   was   measured   by   an   $ac$  technique
\cite{sullivan68} allowing us to measure small samples (here
$\sim  100$ nano-grammes). Heat  was supplied  to the  sample at
a frequency  $\omega$ of  the order  of $70$  Hz by  a light
emitting diode via an optical fiber. The induced temperature
oscillations   were    measured   by   a    $12   \mu$m-diam
chromel-constantan  thermocouple  that   was  calibrated  in
situ.\\
\section{$H-T$ phase diagram}
Fig. 1  displays the temperature dependence  of the specific
heat   at  various   magnetic  fields   up  to   7 T  for
$H||ab$-planes  and up  to 2  T for  $H$ perpendicular  to
those  planes. In zero field, the specific heat discontinuity at $T_c$, $\Delta C(T_c)$,
represents only about 3 $\%$ of the total signal, i.e. 5 to 6 times less than 
in the best polycristalline samples \cite{bouquet}. However, our samples 
are extremely small and the addenda
contribution are thus very large although impossible to estimate 
accurately. 
The data are therefore given in arbitrary
units and the temperature range is limited to  $T
>$ 12 K. No significant increase of 
the transition width, $\Delta T \sim$ 2 K, is observed
for $\mu_0H||c <$ 2 T and $\mu_0H||ab <$ 5 T. Despite a clear broadening above these
chareristic fields, the specific  heat anomaly  remains well  defined in  the 
entire explored  temperature  range. These results imply that the
broadening of the transition width observed in polycristalline samples
\cite{bouquet,park} is due to randomly oriented anisotropic grains
rather than to thermodynamic fluctuations \cite{park}.
The  curves  are  remarkably
identical for both directions once the field is renormalized
by  the  temperature  dependent  anisotropy coefficient (see
inset  of Fig.  1 for  instance). It  is also worth noticing
that  the amplitude  of the  specific heat  jump $\Delta  C$
decreases  much  too  rapidly  with  $H$  implying  a  rapid
increase   of  the   Sommerfeld  coefficient   $\gamma$  and
therefore of the density  of states. An entropy conservation
construction shows that $\gamma  \sim$ 0.9$\gamma _N$ (where
$\gamma_N$ is related to the normal state density of states)
for $H  \sim $0.5 $H_{c2}$  in good agreement  with recent low
temperature measurments  by Bouquet et  al. \cite{bouquet2}.
\\
\begin{figure}[tbp]     \centerline{      \epsfxsize     7cm
\epsffile{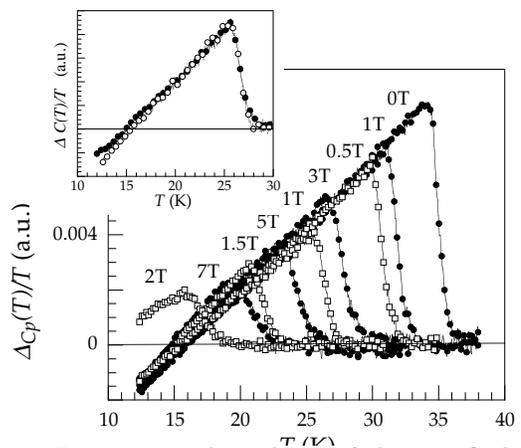}} \caption{Temperature  dependence of the
specific   heat    at   different   magnetic    fields   for
$H||ab-$planes  (closed  circles)  and  $H_{\bot  ab}$ (open
squares).  In  the  inset  :  specific  heat  transition  at
$\mu_0H_{\bot ab}  = $1T and $\mu_0H_{||ab}  = \Gamma.\mu_0H\bot ab \simeq$
3.5  T  showing  that  the  same  curve  is obtained in both
direction  when the  field  is  renormalized by  $\Gamma$. }
\end{figure}
The  critical  temperature  has  been  determined  using the
classical    entropy    conservation    construction.    The
corresponding  $H_{c2}$ values  as well  as the  one deduced
from susceptibility (onset of  the diamagnetic response) and
transport measurments  have been reported in  Fig. 2. In the
latter case,  the critical field  $H_{c2}(T)$ or temperature
$T_{c}(H)$ values  have been defined at  the onset of finite
resistivity (see Fig. 3).\\
Some of  us have previously  shown that the  onset of finite
resistivity coincides with the location of the specific heat
anomaly for $H\bot ab$. Here we show that the same holds for
parallel fields  $H||ab$. Note that  the resistivity reaches
the normal state value at  a magnetic field $H_{R_N}$ higher
than the as-deduced upper  critical field (see Fig.3). Below
we argue that $H_{R_N}$ is due to the surface effects.\\
\begin{figure}[tbp]     \centerline{      \epsfxsize     7cm
\epsffile{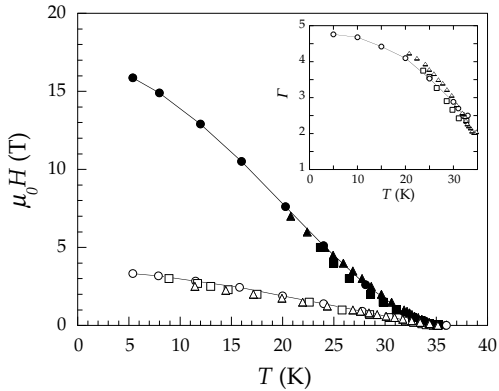}}  \caption{$H-T$  phase  diagram  of the
MgB$_2$  single crystal.  Circles -  $H_{c2}$ from transport
measurments.  Squares -  $H_{c2}$ from  $ac$-susceptibility.
Triangles  - $H_{c2}$  from specific  heat. In  the insert :
temperature dependence of the anisotropy.} \end{figure}
One can see that all three presented methods reveal the same
results in the common range  of applied fields ($\mu _0H \le
$5 T). The  upper critical field perpendicular  to the basal
plane  has a  typical temperature  dependence for  a type-II
superconductor  with  a  linear  shape  near  the zero-field
transition  temperature  and  a  saturation  at  the  lowest
temperatures with $\mu _0H_{c2\bot ab} \simeq 3.5$ T. On the
other hand the parallel upper critical field has a different
strength and  shape: close to  $T_c$, $H_{c2||ab} $  reveals
a positive  curvature  which  changes  to  negative below 20
K and  saturates   to  about  $17$   Tesla  at  the   lowest
temperatures. A direct consequence of the different forms of
the temperature dependencies of these two critical fields is
a temperature  dependent anisotropy  factor $\Gamma$.  Then,
$\Gamma \sim $5 is found at low temperatures but it is about
2 near $T_c$ (see inset of Fig. 2).\\
The presented  data extend the  ones previously obtained  by
some of  us for $\mu_0H  < 7$ T  on crystals from  the same batch
\cite{welp}.  It  has  been   suggested  that  the  positive
curvature observed here for $H||ab$  is a consequence of the
two-gap  structure  \cite{shulga}.   However,  it  is  worth
mentioning  that  a  very  similar  behavior  has  also been
observed in conventional  superconductors including NbSe$_2$
\cite{woollam74},   borocarbides   \cite{lipp02}   and   was
believed  to  be  a   universal  characteristic  of  layered
compounds \cite{woollam74}.  This is even  more "general" as
it  has also  been observed  in the  isotropic (K,Ba)BiO$_3$
system  \cite{blanchard02}. The  origin of  this effect thus
still has to be clarified.\\
\begin{figure}[tbp]     \centerline{      \epsfxsize     8cm
\epsffile{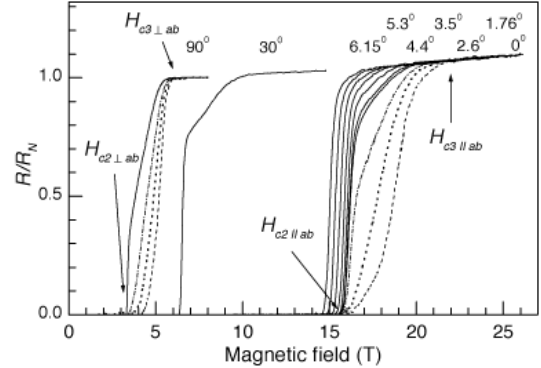}}  \caption{Magnetic field  dependence of
the  MgB$_2$ resistance  at $T  = 5.4$  K for various angles
between  the field  and the  $ab-$plane. The  critical fieds
$H_{c2}$  and  $H_{c3}$  are  marked  by  the arrows for two
principal  field  orientations.  For  these  fields also the
effect of  current density is  presented (dashed lines  from
right to left : $j = $17,  50, 170 A/cm$^2$,  solid lines $j = 
$500  A/cm$^2$). }
\end{figure}
Fig.  3  represents  the   resistive  transitions  from  the
superconducting to normal state taken at 5.4 K for different
angles between  $H$ and the $ab$-planes  of the crystal. The
measurments  have  been  done  at  the  current  density 500
A/cm$^2$  (solid  lines).  The  current  dependence  of  the
transition for perpendicular fields  ($H\bot ab$) - which is
also  highly  non-ohmic  -  has  been  attributed to surface
conductivity in a previous paper  by some of us \cite{welp}.
We  have  found  similar,  albeit  a bit weaker\cite{welp1},
effects  for parallel  fields ($H||ab$)  down to  the lowest
temperatures.   In   Fig.   3   for   both  principal  field
orientations  $H||ab$  and  $H\bot  ab$  an  effect  of  the
measuring   current  is   presented.  It   shows  that   the
transitions  are  indeed   current  dependent.  For  current
densities  equal to  17,  50,  170 A/cm$^2$  transitions are
smooth  and   for  $H\bot  ab$  the   onset  of  the  finite
resistivity  is  even  shifted   towards  higher  fields  for
decreasing current densities. For $j  \ge $ 500 A/cm$^2$ the
onset of the transition is  not changed anymore (measured up
to 2000  A/cm$^2$) and shows  up as an  abrupt jump at  $H =
H_{c2}$ (for  $H||ab$ already at  170 A/cm$^2$) followed  by
a smooth  transition  towards  the  normal  state resistance
$R_N$ . The ratio between the  fields at the onset of finite
resistivity at 500 A/cm$^2$ ($H_{R=0} = H_{c2}$) and the one
at the end of the  transition at the lowest current $H_{R_N}
= H_{c3}$ is about  1.8 for $H\bot  ab$ and 1.4  for $H ||  ab$.
This ratio remains constant for all temperatures apart those
very close to  $T_c$, where it is affected  by the intrinsic
width of the transition $\Delta T_c$. \\
Our  measurments can  be compared  to the  ones obtained  in
a pioneering   work   on    surface   effects   in   type-II
superconductors by Hempstead  and Kim \cite{hempstead}. They
performed    measurments    on    Nb$_{0.5}$Ta$_{0.5}$   and
Pb$_{0.83}$In$_{0.17}$ sheets. For  fields parallel to the
surface of the sheets (that is a favorable configuration for
appearance  of  surface   superconductivity)  they  obtained
results strikingly  similar to ours. For sufficiently low current densities,
the onset of  a finite
resistance is sensitive to the excitation current since the
surface  sheath  could  bear  a  current  keeping  the  zero
resistance  even   above  $H_{c2}$.  However, above  some  threshold
current, this surface superconductivity is destroyed and 
the onset field does not change anymore; it has thus been
identified  with  the  upper  critical  field  $H_{c2}$ \cite{hempstead}. 
As obverved in our data, for
$H = H_{c2}$ the  resistance of the sheets increased sharply to  a fraction of
$R_N$  which was  bigger for  higher current  and then,  the
resistance  finally went   smoothly  to  the   normal  state  $R_N$.
$H_{c3}$  was  best  defined  when  measured  at  the lowest
possible  current  density,  but the experimental $H_{c3}/H_{c2}$
ratio scattered  between 1.6 and  1.96 and droped   down   to  1.14 
for  copper coated Pb-In   sheets.   Indeed,  the
configuration  of our  transport experiment  with $H||ab$ is
very favorable for surface effects  which must then be taken
into account.  On the other  hand, surface superconductivity
effects should be negligible for $H\bot ab$. However, in our
experimental set-up the current and voltage electrodes overlapping
the  sample  go  over  the  vertical  side planes which thus
inevitably   contribute  to   surface  superconductivity  as
well.\\
$H_{c2}$ in  MgB$_2$ crystals has  been measured by  several
authors in a limited  temperature range. The $H_{c2\bot ab}$
values  of Angst  et  al.  \cite{angst} obtained  via torque
magnetometry  are  in  a  good  agreement  with  ours  while
$H_{c2||ab}$ are slightly higher (implying a slightly higher
anisotropy  parameter).  Similar  $H_{c2}$  values have also
been obtained by Sologubenko  et al. \cite{sologubenko} from
the  magnetic  field  and  temperature  dependencies  of the
thermal  conductivity  below  $6$  Tesla.  They  also  found
a standard dependence for $H_{c2\bot ab}(T)$, and a positive
curvature for $H_{c2||ab}(T)$ above $30$  K. Our data are in
very good  agreement with those obtained  by the measurments
of  the  magnetic  moment  in  MgB$_2$  single  crystals  by
Zehetmayer  et  al.  \cite{zehetmayer}.  Finally, Bud'ko and
Canfield \cite{budko} have arrived to similar conclusions by
extracting   the   superconducting   anisotropy   from   the
magnetization   measurments  on   randomly  oriented  powder
samples. It  was however of  fundamental importance to  show
that   the  "critical   fields"  obtained   from  all  those
measurments  coincide  with   the  thermodynamic  transition
deduced from specific heat measurments.
In  many novel  superconductors fluctuations broadens
the transitions from the superconducting to the normal state
in magnetic field. One of the most striking phenomenon which
has  been  observed  in  these  systems  is the existence of
a melting line $T_m(H)$ above which the vortex lattice melts
into  a  liquid  of  entangled  lines.  The presence of this
liquid  phase  complicates  the  determination  of the upper
critical field $H_{c2}$. For instance, the onset of a finite
resistance  in  those  systems  is  indicating  the  melting
transition (at $H_{R=0}$) while  the upper critical field is
located  at   a  very  end   of  the  resistive   transition
$R(H_{c2})  =  R_N$  (see  for  instance  \cite{samuely98}).
However,   fluctuations  are   not  expected   to  play  any
significant role  in MgB$_2$ as  confirmed by specific  heat
measurments:  the   width  $\Delta  C$   does  not  increase
significantly  with magnetic  field and  the upper  critical
field  occurs  near  the   onset  of  finite  resistance  at
$H_{R=0}$.  A   non  correct  criterion   for  the  $H_{c2}$
determination partly explains the  discrepancy in the values
deduced from transport measurments \cite{pradhan,eltsev}.\\
\begin{figure}[tbp]     \centerline{      \epsfxsize     7cm
\epsffile{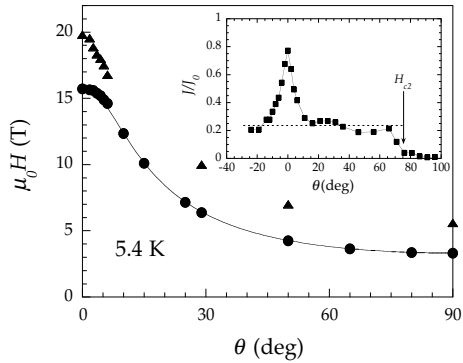}}  \caption{Angular   dependence  of  the
upper critical  field (open circles) at  5.4 K together with
the  fits  by  the  Lawrence-Doniach  formula. Triangles are
"critical  fields"  taken  at  the  end  of  the resistivive
transition $H_{R_N}$ showing the cusp-like behavior at small
angles.  In the  inset: angular  dependence of  the critical
current  at  $\mu_0H =$2  T  and   $T=18$  K  ($J_0  \sim $ 5x10$^3$
A/cm$^{-3}$).} \end{figure}
Next,  we discuss  the  angular  dependence of  the critical
field.  As shown  in Fig.  3, the  resistivity rapidly drops
towards zero at $\mu_0H \sim$  16 T for $H||ab$ and progressively
shifts  towards lower  fields  as  the angle  increases. The
corresponding $H_{c2}(\Theta)$ can be well fitted by the the
simple                    ellipsoidal                   form
$(\frac{H_{c2}(\Theta)\sin{\Theta}}{H_{c2\bot       ab}})^2+
(\frac{H_{c2}(\Theta)\cos{\Theta}}{H_{c2||ab}})^2  =  1$  of
Lawrence    and    Doniach    \cite{LD}    for   anisotropic
three-dimensional superconductors  with $\Gamma =  4.8$ (see
Fig.4). A similar dependence has  been observed at 26 K (not
shown) but  with $\Gamma =$  3.4. By triangles  we have also
plotted the characteristic fields $H_{R_N}$. If surface 
conductivity becomes important for  $H  \bot  ab$  and  $H||ab$  
($H_{R_N}$ is then close  to  $H_{c3}$), this effect is diminished for other field
orientations and the $H_{R_N}/H_{c2}$ ratio  
drops down to $\approx 1.1$. Surface conductivity effects can 
thus account for the cusp-like
behavior of  $H_{R_N}$ at small angles  which has been first
attributed     to     $H_{c2}$      by     other     authors
\cite{eltsev,ferdeghini}.\\
\section{Peak effect and angular  dependence of the critical
current}
Another interesting phenomenon is  the observation of a peak
effect in the critical current.  This effect shows up in our
transport measurments  in the same way  as in\cite{welp} and
is  also  visible  in  our  susceptibility  data  in a small
magnetic field range $\sim 2-3$  T for $H$ perpendicular
to  the $ab-$plane.  This increase  in the  critical current
appears as  a dip in the  susceptibility in both temperature
and magnetic field sweeps as also reported by Pissas et al.
for the same field orientation \cite{pissas}. Surprizingly,
the peak effect  is only visible in a  narrow magnetic field
range. For  lower magnetic fields,  the peak is  replaced by
a rapid drop of the susceptibility and it has been suggested
by  Pissas  et  al.  \cite{pissas}  that  this drop could be
associated  with  a  first  order  transition  in the vortex
lattice  in  analogy  with  high  $T_c$  oxides. However, in
MgB$_2$, the  peak effects appears  close the $H_{c2}$  line
and no  sign for the existence  of a vortex liquid  could be
obtained in  our experiments. \\
\begin{figure}[tbp]     \centerline{      \epsfxsize     7cm
\epsffile{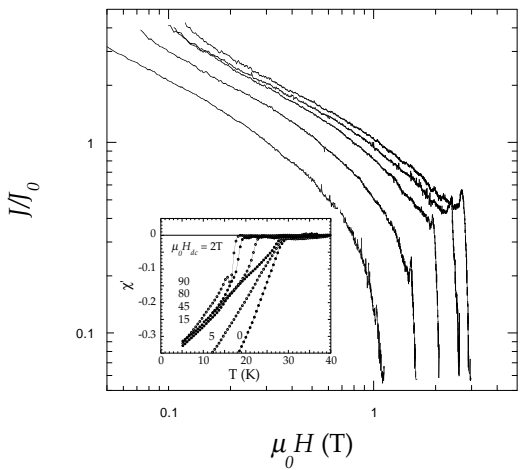}} \caption{ Magnetic  field dependence of
the critical current at $H\bot ab$  at $T$(K) = 8.5, 13, 17,
21.5 and  26 (from top to  bottom) showing a peak  effect at
low temperatures  ($J_0 \approx$ 5 10$^3$  A/cm$^3$). In the
inset : temperature dependence  of the $ac$-suscetibility at
$\mu_0H  = $2  T  for  different  angles  between  $H$  and  the
$ab-$plane.} \end{figure}
 The  peak is here reflecting
the  fact that  the shear  energy of  the flux lattice drops
towards  zero more  rapidly than  the pinning  energy in the
vicinity of  the normal state allowing  to the vortex matter
to accomodate more efficiently to  the disorder close to the
transition \cite{pipard}. Moreover, as  shown in Fig. 5, the
value of  the peak current rapidly  decreases for increasing
temperature  (i.e.   for  decreasing  magnetic   field)  but
probably  still exists  at high  temperature. Similarly,  as
shown in the inset of Fig.  5, the dip in the susceptibility
(that is the peak in the current density)
also  rapidly disappears  as the  magnetic fields  is turned
away from  the $c-$axis. However,  the peak effect  is still
visible   in   transport   measurments   for  $H||ab$  again
suggesting  that  the  peak  has  shifted  towards lower $J$
values when  the magnetic field  was turned and  thus disappeared
from our  experimental window. The  $J$ values presented  in
Fig.5 have  been deduced from the  magnetic field dependence
of  the  $ac-$susceptibility.  Indeed,  in  the  non  linear
regime, the $ac$ response is  related to the current density
$J$ through : $ \chi(h_{ac}, \omega, T) = F(h_{ac}/J(\omega,
T))$  where $F(x)$  depends  on  the pinning  mechanism, the
sample  geometry  and  the   flux  dynamics.  $F$  has  been
determined experimentally following the procedure introduced
by Pasquini {\it et al.} \cite{pasquini} and is in very good
agreement  with  the  numerical  calculations  performed  by
Brandt \cite{Brandt} for bulk pinning in cylinders (with the
appropriate thickness/radius ratio).  Note that our critical
current  values are  much  smaller  than the  one previously
obtained  in polycrystals  \cite{wen} but,  as shown  below,
$J$ sharply increases when the field  is applied parallel to the
$ab-$ planes.\\
As  shown in  the inset  of Fig.5,  at low  temperature, the
susceptibility  (and thus the critical  current) only weakly
depends on $\theta$ down to $\theta \sim  10$ degrees and
the rapid  drop of the  susceptibility close to  $H_{c2}$ is
progressively   shifted  towards   higher  temperature   for
decreasing $\theta$ values. Finally  close to the $ab-$plane
the  susceptibility  rapidly   decreases  showing  that  the
pinning of the vortices is  much more efficient for $H||ab$.
The corresponding $J/J_0$ values at  $18$ K are displayed in
the inset of  Fig. 4 showing that $J$  increases by a factor
of about $4$ for $H$  parallel to the $ab-$planes. A similar
behavior has  been observed in the  critical current deduced
from transport  measurments in $c-$axis  oriented thin films
\cite{sen}  and Eltsev  {\it  et  al.} recently  reported on
a rapid  drop of  the dissipation  when the  magnetic field is
applied  parallel  to  the  $ab-$planes  in  single crystals
\cite{eltsev}. A  similar cusp in  the critical current  has
been observed  in high $T_c$ cuprates  in which the vortices
are  strongly pinning  by the  weakly superconducting layers
between  the  CuO$_2$  planes  when  the  field  is  applied
parallel  to  those  planes  (so  called  intrinsic  pinning
\cite{tachiki}). However,  the origin of  the strong pinning
of the $ab-$  planes in diborides still has  to be clarified
given the rather small anisotropy of this material.\\
\section{CONCLUSION}
In summary,  the upper critical field  has been deduced from
specific   heat,   high   magnetic   field   transport   and
$ac$-susceptibility   measurments  for   $H$  parallel   and
perpendicular to the  $ab-$planes. $H_{c2}$ perpendicular to
the  planes  reveals  a  conventional temperature dependence
with $\mu_0H_{c2\bot  ab}(0) \simeq $ 3.5  Tesla but the parallel
critical  field  with  $\mu_0H_{c2||ab}(0)  \simeq$  17 Tesla has
a positive   curvature   at   temperatures   above   20   K.
Consequently,    the    anisotropy    factor    $\Gamma    =
H_{c2||ab}/H_{c2\bot    ab}$   is    temperature   dependent
decreasing from  $\sim 5$ at  low temperature towards  $\sim
2$ close  to $T_c$. The critical  current deduced from the
susceptibility measurments  present a sharp  peak effect for
$\mu_0H\bot  ab  \sim  2-3$  T.   At  low  temperatures,  $J$ only depends
weakly on the angle between  $H$ and the $ab-$plane for
$\theta > 10$ degrees but rapidly increases for $H||ab$.\

\end{multicols}

\begin{references}
\bibitem{nagamatsu}
J.  Nagamatsu et al., Nature {\bf 410}, 63 (2001).
\bibitem{liu}
A. Y. Liu et al., Phys. Rev. Lett. {\bf 87}, 087005 (2001).
\bibitem{choi}
H. J. Choi et al., Phys.  Rev. B {\bf 66}, 020513(R) (2002);
Nature {\bf 418}, 758 (2002).
\bibitem{bouquet}
F. Bouquet et al., Phys. Rev. Lett. {\bf 87}, 047001 (2001).
\bibitem{szabo}
P. Szab\'o et al., Phys. Rev. Lett. {\bf 87}, 137005 (2001).
\bibitem{kim}
K. H. P. Kim et al., Phys. Rev. B {\bf 65}, 100510(R), 2002.
\bibitem{sullivan68}
P.F. Sullivan and G. Seidel \rm Phys. Rev. {\bf 173}, 679 (1968).
\bibitem{park} 
T. park et al., cond-mat/0204233.
\bibitem{bouquet2}
F. Bouquet et al., cond-mat/0207141.
\bibitem{welp}
U. Welp et al., cond-mat/0203337.
\bibitem{shulga}
S. V. Shulga et al., cond-mat/0103154.
\bibitem{woollam74}
J.A.Woollam, R.Somoano and P.O'Connor Phys. Rev. Lett. {\bf 32}, 712 (1974).
\bibitem{lipp02}
D. Lipp et al., Europhys. Lett. {\bf 58}, 435 (2002) and references therein.
\bibitem{blanchard02}
S.Blanchard et al., Phys. Rev. Lett. {\bf 88}, 177201 (2002).
\bibitem{welp1}
Ref.\cite{welp}   has discussed  possible  arguments   for  weaker
surface  superconductivity effects in the $H||ab$ configuration
based on surface electronic states in MgB$_2$.
\bibitem{hempstead}  C. F.  Hempstead and  Y. B.  Kim, Phys.
Rev. Lett. {\bf 12}, 145 (1964).
\bibitem{angst}
M. Angst et al., Phys. Rev. Lett. {\bf 88}, 167004 (2002).
\bibitem{sologubenko}
A. V. Sologubenko et al., Phys. Rev. B {\bf 65}, 180505(R) (2002).
\bibitem{zehetmayer}
M. Zehetmayer et  al., cond-mat/0204199.
\bibitem{budko}
S. L. Bud'ko and P. C. Canfield, cond-mat/0201085.
\bibitem{samuely98}
P. Samuely et al.,\rm Europhys. Lett. {\bf 41}, 207 (1998).
\bibitem{pradhan}
A. K. Pradhan et  al., Phys. Rev. B {\bf 64}, 212509 (2001).
\bibitem{eltsev}
Yu. Eltsev et al., Phys. Rev. B {\bf 65},140501(R) (2002).
\bibitem{LD}
W. E. Lawrence and S. Doniach, in {\it Proc. 12th Int. Conf.
Low Temp. Phys.}, ed. E. Kanda, Keigaku, Tokyo 1981, p. 361.
\bibitem{ferdeghini}
C.  Ferdeghini  et  al.,  cond-mat/0203246.
\bibitem{pissas}
M. Pissas, S. Lee, A. Yamamoto and S. Tajima, cond-mat/0205561
\bibitem{pipard}
A. B. Pipard, Phil. Mag. {\bf 19}, 217 (1967).
\bibitem{pasquini}
G. Pasquini, L. Civale, H. Lanza  and G. Nieva, Phys. Rev. B
{\bf 59}, 9627 (1999).
\bibitem{Brandt}
E.H. Brandt, Phys. Rev. B {\bf 55}, 14513 (1997) and references therein.
\bibitem{wen}
H. H. Wen, S. L. Li, Z. W. Zhao, H. Jin, Y. M. Ni, Z. A. Ren, G. C.
Che and Z. X. Zhao,
Supercond. Sci. Technol. {\bf 15}, 315 (2002).
\bibitem{sen}
S. Sen, A. Singh, D. K. Aswal, S. K. Gupta, J. V. Yakhmi, V.C. Sahni, E.
M. Choi, H. J. Kim, K. H. P. Kim, H. S. Lee, W. N. Kang and S. I. Lee,
Phys. Rev. B {\bf 65}, 214521 (2002).
\bibitem{tachiki}
M. Tachiki and S. Takahaski, Solid State Com. {\bf 72}, 1083 (1989).
\end{references}
\end{document}